\RequirePackage{fix-cm}
\documentclass[twocolumn]{svjour3}          
\smartqed  
\usepackage{epsfig,amsmath,amssymb,graphics,amsfonts,psfrag}
\usepackage[letterpaper,margin=1in]{geometry}

\def\R{\mathbb{R}}

\def\N{\mathbb{N}}
\def\P{\mathbb{P}}
\def\E{\mathbb{E}}

\newcommand{\be}{\begin{equation}}
\newcommand{\ee}{\end{equation}}
\newcommand{\bea}{\begin{eqnarray}}
\newcommand{\eea}{\end{eqnarray}}
\newcommand{\beann}{\begin{eqnarray*}}
\newcommand{\eeann}{\end{eqnarray*}}
\newcommand{\benn}{\begin{equation*}}
\newcommand{\eenn}{\end{equation*}}

\def\ra{\rightarrow}
\def\I{\infty}
\def\I{\infty}

\newcommand{\cI}{{\mathcal I}}  
\newcommand{\cN}{{\mathcal N}}  
\newcommand{\cO}{{\mathcal O}}  

\begin{document}

\author{Christian Kuehn}

\institute{C. Kuehn \at
              Vienna University of Technology,\\ 
              Institute for Analysis and Scientific Computing, \\
	      Vienna, 1040, Austria.\\
              \email{ck274@cornell.edu}  
}

\journalname{arXiv preprint}
 
\title{Warning Signs for Wave Speed Transitions\\ of Noisy Fisher-KPP Invasion Fronts}


\maketitle

\begin{abstract}
Invasion waves are a fundamental building block of theoretical ecology. In this study we aim to take the first steps to
link propagation failure and fast acceleration of traveling waves to critical transitions (or tipping points). The approach
is based upon a detailed numerical study of various versions of the Fisher-Kolmogorov-Petrovskii-Piscounov (FKPP) equation.
The main motivation of this work is to contribute to the following question: how much information do statistics, collected
by a stationary observer, contain about the speed and bifurcations of traveling waves? We suggest warning signs based upon 
closeness to carrying capacity, second-order moments and transients of localized initial invasions. 
\end{abstract}


\section{Introduction}  
\label{sec:intro}

The propagation of waves has been a central topic in spatial ecology for a long time. A primary motivation arises from 
fronts where a new species is introduced into an environment or an existing species considerably extends its habit. A classical 
example is the spread of muskrats in central Europe \cite{Skellam}. Other documented examples are butterflies and bush crickets in the UK \cite{Thomasetal}
and the cane toad invasion in Australia \cite{PhilippsBrownWebbShine}. Also bacterial growth 
\cite{MimuraSakaguchiMatsushita,GoldingKozlovskyCohenBen-Jacob} shows very similar spreading and wave phenomena. The references 
in \cite{Hastingsetal,SimberloffGibbons} contain even more examples. 

From a theoretical perspective a first groundbreaking result is the modelling of invasion waves via reaction-diffusion equations by 
\cite{Fisher} and Kolmogorov, Petrovskii, Piscounov \cite{KolmogorovPetrovskiiPiscounov} (FKPP) who studied the partial differential equation 
(PDE) 
\be
\label{eq:base_PDE}
\frac{\partial u}{\partial t}=\frac{\partial^2u}{\partial x^2}+u(1-u).
\ee
There are many different aspects that could be included in a reaction-diffusion model which 
are very interesting to match theory and experiment; see \cite{Hastingsetal,Hilkeretal,FaganLewisNeubertvandenDriessche,MetzMollisonvandenBosch}. 
Nevertheless, the basic guiding principles obtained from simple models are still highly relevant. Here we shall restrict
ourselves to the study of the following stochastic partial differential equation (SPDE)
\be
\label{eq:base_SPDE}
\frac{\partial u}{\partial t}=\frac{\partial^2u}{\partial x^2}+f(u)+\text{'noise'}.
\ee 
For now, the reader may just think of the classical FKPP nonlinearity $f(u)=u(1-u)$ and some noise process that vanishes at
zero-population level; for more technical details see Sections \ref{sec:background_FKPP}-\ref{sec:background_SPDE}. The detailed choices 
are discussed later. 

The main theme of this paper is the interplay between invasion waves and so-called critical transitions 
\cite{Schefferetal,KuehnCT1}. Basically, critical transitions (or tipping points) are drastic sudden changes in dynamical systems; for 
some background and details see Section \ref{sec:background_CT}. The first major question is whether \eqref{eq:base_PDE}-\eqref{eq:base_SPDE}
can undergo a 'critical transition'. We discuss this question from a more technical perspective in Section \ref{sec:background_CT}. On
a heuristic level, one may just consider a parameter in \eqref{eq:base_SPDE} that is slowly varying. Suppose that there exists a wave 
with positive speed for some parameter range while the wave is stationary (or reverses direction) for another parameter range. Whether an invasion 
reaches a new habitat or not can have drastically different consequences so one probably would like to refer to this situation as a 
critical transition. 

Another case we shall consider in this paper is the situation where the wave speed becomes infinite at a special 
parameter value. Hence, a small parameter variation can cause a dramatically accelerating invasion wave.\\  

The next step is to check whether early-warning signs for a critical transition exist. In this context, changes in vegetation patterns 
have been the main motivation recently \cite{Kefietal,HirotaHolmgrenvanNesScheffer}. There are only a few studies on early-warning signs 
for spatial systems \cite{Dakosetal1,Dakosetal2}. In fact, early-warning signs for noisy waves generated by SPDEs have not been considered 
yet. This paper makes a first step in this direction. We focus on transitions for the wave speed ({e.g.}~propagation failure) 
as it controls when and where an invasion front appears. Although some detailed measurements of waves are available
\cite{Holway,Lonsdale} it is very difficult to obtain precise global empirical information \cite[p.92]{Hastingsetal} about a wave. Here
we restrict ourselves to a single spatial observation location {i.e.}~records by a single stationary 'ecological observer' over a fixed 
time interval. The general idea that one may obtain spatial conclusions from local observations is not new \cite{FilipeOttenGibsonGilligan}.
However, our detailed comparative numerical study of several different variants of \eqref{eq:base_SPDE} with a focus on local early-warning
signs seems to be a completely new direction; for more details on the numerical methods see Section \ref{sec:background_numerics}. The main
themes and results from the numerical studies are the following:

\begin{itemize}
 \item[(a)] A description of the statistics for early-warning signs in SPDEs with wave propagation failure based upon closeness to carrying capacity and 
second-order moments.
 \item[(b)] A comparative study of (a) for different noise types (white, space-time white) and different multiplicative noise nonlinearities (parametric, 
finite-system size, etc.).
 \item[(c)] Investigation of statistics near continuous wave speed transitions (and their unpredictability) for Allee effect nonlinearities in the 
deterministic part of the FKPP SPDE.
 \item[(d)] Suggestion of transient minima to analyze wave propagation failure and wave speed blow-up.  
\end{itemize}

Beyond the technical contributions we also try to link different methodologies. We combine approaches from biological invasions, critical transitions, 
Fisher-KPP (and Nagumo) waves, SPDEs and numerical methods. This approach should also be helpful to link several, mostly distinct, communities such as 
theoretical ecology, waves in theoretical physics and mathematical methods for SPDEs. 

The paper is organized as follows. In Sections \ref{sec:background_CT}-\ref{sec:background_numerics} we give brief reviews of the essential 
facts required for the remaining part of the paper. Due to the interdisciplinary aspects, the brief reviews seem necessary. Readers familiar with all the 
background may forward to Section \ref{sec:lin_noise} where the multiplicative noise case for the FKPP SPDE and statistical warning signs are studied. 
The nonlinear noise case is considered in Section \ref{sec:nonlin_noise} and the Allee effect in Section \ref{sec:Allee}. Section \ref{sec:noncompact} on 
transient phenomena and the influence of initial conditions concludes the main part of the paper. In Section
\ref{sec:outlook} a number of generalizations and open problems are listed. 

\section{Background - Critical Transitions}
\label{sec:background_CT}

A primary motivation to study critical transitions (or tipping points) arose from ecology, 
{e.g.}~due to the theoretical work of Scheffer and co-workers \cite{SchefferCarpenter,Schefferetal1,ScheffervanNes}. Then it became clear 
from many distinct applied problems \cite{Schefferetal} as well as from abstract mathematical considerations \cite{KuehnCT1,KuehnCT2} 
that many features for early-warning signs are generic across many dynamical systems. Recent studies of laboratory \cite{DrakeGriffen,Veraartetal} and 
full ecosystem \cite{Carpenteretal1} experiments re-inforced this viewpoint. 

Here we recall a few aspects of critical transitions for finite-dimensional systems relevant for this paper. Consider the pitchfork bifurcation 
normal form \cite[p.282]{Kuznetsov}
\be
\label{eq:pitchfork}
\frac{dw}{dt}=w'=\mu w+w^3,\qquad \text{for $w\in\R$, $\mu\in\R$.}
\ee 
The homogeneous trivial branch $\{w=0\}$ consists of stable equilibria for $\mu<0$ and unstable equilibria for $\mu>0$ since the linearized
system around $w=0$ is $W'=\mu W$ with solution $W(t)=W(0)e^{\mu t}$. The bifurcation at $\mu=0$ is sub-critical with two unstable branches 
$\{w=\pm\sqrt{-\mu}\}$ for $\mu<0$. 

Consider a slow parameter variation $\mu'=\epsilon$ with $0<\epsilon\ll1$ and $\mu(0)<0$. Orbits near the homogeneous 
branch will reach a neighborhood of $(w,\mu)=(0,0)$ and then jump away quickly indicating a critical transition \cite[Fig.3(c)]{KuehnCT1}.
Before the jump the system is slow to recover from perturbations ('slowing down') for $\mu<0$ since $W(0)e^{\mu t}\ra W(0)$ as $\mu\ra 0$ for 
fixed $t$. For a deterministic system, it is impossible to measure the slowing-down effect once it starts tracking the homogeneous branch
$\{w=0\}$ {i.e.}~it is exponentially close to $w=0$. However, for a stochastic version of \eqref{eq:pitchfork} given by  
\be
\label{eq:pitchfork1}
w'=\mu w+w^3+\text{'noise'}
\ee  
the random perturbations can constantly kick the system away from the trivial branch. Extracting statistics from these perturbations 
can make the slowing down effect measurable \cite{Schefferetal,KuehnCT1}. This is one motivation to study stochastic traveling waves 
\eqref{eq:base_SPDE}.\\

An important question is which bifurcation points or quantitative transitions we would like to classify as critical transitions. In 
multiple time scale systems, such as \eqref{eq:pitchfork1} augmented with $\mu'=\epsilon$, the classification of local bifurcation points 
is relatively straightforward \cite[Sec.2-3]{KuehnCT2}. The mathematical classification and early-warning signs from \cite{KuehnCT1,KuehnCT2}
can be applied to many pattern-forming bifurcations in spatially extended systems on bounded domains. One first derives the amplitude equations
on the domain locally \cite{CrossHohenberg}. Only a discrete set of eigenvalues occurs \cite[p.210]{Hoyle} and the usual local bifurcations
for a finite number of eigenvalues passing through the imaginary axis can often be applied. 

For patterns on unbounded domains the situation is less clear. We do not offer any solution to this problem and consider 
an example to illustrate the difficulties. Consider a traveling wave solution $u(x,t)=u(x-ct)$, {e.g.}~for \eqref{eq:base_PDE}, with 
$(x,t)\in \R\times \R^+$ which satisfies 
$u(x,0)=1$ for $x\leq0$ and $u(x,0)=0$ for $x> 0$. Imagine a habitat $[x_1,x_2]\subset \R$ with $x_{1,2}>0$ and define the mapping
\benn
\cI(u,T)=\frac{1}{|x_2-x_1|}\int_{x_1}^{x_2} |u(x,T)|dx.
\eenn 
If the invasion wave spreads towards $x=\I$ ($s>0$) and saturates at the carrying capacity $u\equiv 1$ then there exists a finite time $T_i$ such that
for all $T\geq T_i$ we have $\cI(u,T)=1$. If a slow parameter variation causes the wave to become stationary $(s=0)$ then $\cI(u,T)=0$ for all $T\geq 0$.
Although this indicates how one may define one possible critical transition scenario for waves, the situation is actually unclear since for fixed 
$T>0$ one may have $\cI(u,T)=0$ for $s>0$ and $s=0$. This illustrates again that global definitions are intricate \cite[Sec.8]{KuehnCT2}.    

For this paper we simply rely on the intuitive notion that the transition to a standing wave and also the transition to wave speed blow-up 
are important in the context of critical transitions and early-warning signs.

\section{Background - FKPP Equation(s)}
\label{sec:background_FKPP}
 
A more general version of the PDE \eqref{eq:base_PDE} studied by Fisher \cite{Fisher} as well as by Kolmogorov, Petrovskii and 
Piscounov \cite{KolmogorovPetrovskiiPiscounov} is given by
\be
\label{eq:FKPP_det}
\frac{\partial u}{\partial t}=D\frac{\partial^2u}{\partial x^2}+f(u;\mu)
\ee 
for $u=u(x,t)$, $(x,t)\in\R\times [0,\I)$. The parameter $D>0$ controls the diffusion and if $f(u;\mu)=\mu f(u)$ 
then $\mu>0$ can be interpreted as a growth rate. Of course, an initial condition has to be specified. 
Often one considers $u(x,t=0)$ either with compact support localized near $x=0$ or an initial condition with 
Gaussian decay. The localized initial condition for a population $u$ to appear in a new environment is not 
only a mathematical simplification but does occur under realistic conditions {e.g.}~due to global 
long-range transportation networks \cite{KoelzschBlasius}.

The nonlinearity $f:\R^2 \ra \R$ represents growth and saturation effects and is required to satisfy 
the conditions
\benn
\begin{array}{lcllcl}
f(0;\mu)&=&0,\qquad & f'(0;\mu)&>&0,\\
f(1;\mu)&=&0,\qquad & f'(1;\mu)&<&0.\\
\end{array}
\eenn
The classical example is logistic growth $f(u;\mu)=\mu u(1-u)$. In this case 
one may rescale $t\mapsto t/\mu$, $x\mapsto x\sqrt{D/\mu}$ to obtain from \eqref{eq:FKPP_det} the FKPP equation
\be
\label{eq:FKPP_det1}
\frac{\partial u}{\partial t}=\frac{\partial^2u}{\partial x^2}+u(1-u).
\ee 
Initially, the FKPP equation \eqref{eq:FKPP_det1} modeled the spread of genes in a population 
but it has since become a paradigmatic model for populations dispersing under the influence of 
diffusion \cite[p.439-444]{Murray1}. Using a traveling wave ansatz $u(x,t)=u(x-ct)=:u(\xi)$ for \eqref{eq:FKPP_det1}
yields the ODE
\be
\label{eq:ODE_waveframe}
\frac{d^2 u}{d\xi^2}+c\frac{du}{d\xi} +u(1-u)=0.
\ee
Analyzing \eqref{eq:ODE_waveframe} in the phase space variables $(u,u')=:(u,v)$ shows that the point $(u,v)=(1,0)$ is a saddle and 
$(u,v)=(0,0)$ is a stable node or spiral. It is straightforward \cite[p.441-442]{Murray1} to check that heteroclinic 
orbits from $(1,0)$ to $(0,0)$ with $u\geq0$, which correspond to non-negative traveling waves, can only exist in the 
stable node case for wave speeds $c\geq 2$. Since the orbit is directed from $u=1$ to $u=0$ one also refers to this situation
as the stable state $u\equiv1$ invading the unstable state $u\equiv0$; see Figure \ref{fig:2}(a).\\

\textit{Remark:} Wave speeds $c>0$ correspond to waves traveling to the right. However, the FKPP equation 
\eqref{eq:FKPP_det1} is invariant under the symmetry $x\ra -x$ so that a localized initial condition near $x=0$ triggers 
a pair of fronts, one traveling to the left one to the right.\\ 

It is known that the traveling wave solutions to \eqref{eq:FKPP_det1}
form the important solution set \cite[Thm 1.4-1.5]{HamelNadirashvili}. The minimal wave speed $c^*_{FKPP}=2$ is the asymptotic 
speed of propagation for the FKPP nonlinearity \cite{BenguriaDepassier}. Since the wave speed is determined by the 
linearized problem 
\be
\label{eq:FKPP_det2}
\frac{\partial \tilde{u}}{\partial t}=\frac{\partial^2\tilde{u}}{\partial x^2}+\tilde{u}
\ee
at the leading edge near $(u,v)=(0,0)$ as detailed in \cite[p.38-42]{vanSaarloos} one also refers to the traveling wave
with $c^*_{FKPP}=2$ as a pulled front. For a more general nonlinearity $f(u;\mu)$ pushed fronts can exist where the asymptotic 
wave speed $c$ is larger than the linear spreading speed $c^*$ \cite[p.56]{vanSaarloos}. The wave speed for both types 
is asymptotic and only achieved after a transient period. For pulled fronts the asymptotic
expansion yields \cite[p.78]{vanSaarloos}
\benn
c(t)=c^*-\frac{k_1}{t}+\frac{k_2}{t^{3/2}}+\cO\left(\frac{1}{t^2}\right),\qquad \text{as $t\ra \I$}
\eenn
with explicitly computable positive constants $k_{1,2}>0$. Hence, the wave speed is approached from below by a power law for pulled 
fronts. For pushed fronts the convergence to the asymptotic speed is exponentially fast \cite[p.74]{vanSaarloos}. Another 
correction occurs when a cutoff for the reaction term is introduced \cite{BrunetDerrida} which leads to a logarithmic correction
term. Furthermore, if an initial condition does not decay fast enough as $|x|\ra \I$ then faster speeds than $c^*$ occur 
\cite[p.46]{vanSaarloos}. In particular, for an initial condition decaying like $\cO(e^{-\alpha |x|})$ for $\alpha>0$ the speed increases as 
$c(\alpha)=\cO(1/\alpha)$ as $\alpha \ra 0$ \cite{Roquesetal}.

The results for invasion fronts of the FKPP equation already indicate that the variety of scaling behaviors 
could be ideal to determine early-warnings. In fact, wave spreading in the stochastic case is even more intricate 
\cite{LemarchandLesneMareschal,Lewis}.  

\section{Background - Stochastic PDEs}
\label{sec:background_SPDE}

As a stochastic generalization of \eqref{eq:FKPP_det} the intuitive idea is to consider the equation
\be
\label{eq:eta_SPDE}
\frac{\partial u}{\partial t}=D\frac{\partial^2u}{\partial x^2}+\mu f(u)+g(u)\eta(x,t)
\ee 
where $\eta(x,t)$ formally represents the 'noise'. Here we consider two choices for the term 
$\eta(x,t)$. The simplest is to consider a real-valued (1D) Brownian motion $B(t)$ \cite[Chapter 8]{Durrett2a} with 
mean $\E[B(t)]=0$ and covariance $\E[B(t)B(s)]=\min(t,s)$ for $0\leq s\leq t$. Then 
white noise can be defined via $\eta(x,t)=\eta(t)=\dot{B}$ where the derivative is with respect to time and interpreted in the 
generalized sense \cite[p.52-53]{ArnoldSDEold}. The covariance is $\E[\eta(t)\eta(s)]=\delta(t-s)$ and one may then write 
\eqref{eq:eta_SPDE} in two equivalent forms
\be
\label{eq:1Dnoise}
\begin{array}{lcl}
\frac{\partial u}{\partial t}&=&D\frac{\partial^2u}{\partial x^2}+\mu f(u)+g(u)\dot{B}(t),\\ 
du&=&\left[D\frac{\partial^2u}{\partial x^2}+\mu f(u)\right]dt+g(u)dB.\\
\end{array}
\ee 
The existence and regularity theory of \eqref{eq:1Dnoise} is well understood \cite{Flandoli}. If the noise should 
depend on space and time the theory is substantially more involved. 

One possibility is to consider a Hilbert space $U$ (e.g. $L^2(\R)$) and a symmetric non-negative linear operator $Q$ 
acting on $U$ and define a $U$-valued $Q$-Wiener process $W(t)$. If $Tr(Q)<+\I$ there exists a complete orthonormal system 
$\{f_k\}_{k=1}^\I$ such that
\benn
Qf_k=\lambda_k f_k,\qquad \text{for $k\in\N$}
\eenn
where $\{\lambda_k\}_{k=1}^\I$ is a nonnegative bounded sequence. Then one may use the convergent sequence 
\be
\label{eq:series}
W(t)=\sum_{k=1}^{\I}\sqrt{\lambda_k}B_k(t)f_k  
\ee 
with independent one-dimensional Brownian motions $B_k(t)$ as a definition \cite[p.86-89]{DaPratoZabczyk}. As expected 
one has $\E[W(t)]=0$ and $\E[W(t)W(s)]=\min(t,s)Q$ so that $Q$ can be viewed as the covariance operator. In this case one may
formally write \eqref{eq:eta_SPDE} as looking for $u=u(\cdot,t)$ in the form
\be
\label{eq:infnoise}
\begin{array}{lcl}
\frac{\partial u}{\partial t}&=&D\frac{\partial^2u}{\partial x^2}+\mu f(u)+g(u)\dot{W}\\
du&=&\left[D\frac{\partial^2u}{\partial x^2}+\mu f(u)\right]dt+g(u)dW.\\
\end{array}
\ee 
which can be interpreted in a precise integral form \cite[Section 5.1]{DaPratoZabczyk}. A well-developed existence theory 
for \eqref{eq:infnoise} is available \cite[Thm 7.4, Thm 7.6]{DaPratoZabczyk}.

One is tempted to take $Q=Id$ to mirror the finite-dimensional case to obtain a 'white noise' process. However, $Q=Id$ is 
not of trace-class (since $Tr(Q)=+\I$) and the series \eqref{eq:series} does not converge. However, one may construct a 
cylindrical Wiener process for $Q=Id$ \cite[p.96-99]{DaPratoZabczyk} and characterize space-time white noise as 
$\dot{W}=\eta(x,t)$ which has covariance $\E[\eta(x,t)\eta(y,s)]=\delta(x-y)\delta(t-s)$. The existence and regularity theory 
for space-time white noise is slightly more involved and already leads to problems if $x\in\R^2$ \cite[p.54]{Chow}. Since 
we exclusively restrict to $x\in\R$ these problems do not arise here and the existence theory works \cite[Section 6.1]{Hairer}.\\

\textit{Remark:} Instead of viewing the equation on function spaces one may also consider an
approach \cite{Walsh} where the solution $u=u(x,t)$ is a real-valued random field which is a basically equivalent 
\cite{Jetschke} approach. In this case, one has $\E[W(x,t)W(y,s)]=\min(t,s)\min(x,y)$ and that 
space-time white noise is $\frac{\partial^2}{\partial x\partial t}W(x,t)=\eta(x,t)$.

\section{Background - Numerical SPDEs}
\label{sec:background_numerics}

First, we briefly review basic methods to solve the SPDE \eqref{eq:infnoise} numerically for space-time white noise. The 
case \eqref{eq:1Dnoise} will follow as a special case. A natural first step is to start with a spatial discretization \cite{JentzenKloeden}. 
Consider a finite interval $[x_1,x_{N}]\subset \R$ for some $N>1$ and augment \eqref{eq:infnoise} with zero, reflective or periodic boundary 
conditions. Define $(\Delta x):=(x_N-x_1)/N$ and consider the numerical solution $U_j(t)\approx u(x_1+(j-1)(\Delta x),t)$. Then the 
space-discrete version of \eqref{eq:infnoise} is a system of stochastic ordinary differential equations (SODEs)
\be
\label{eq:SODE}
\begin{array}{lcl}
dU_j&=&\underbrace{\left[\frac{D}{(\Delta x)^2}\sum_{l=1}^NL_{jl}U_l+\mu f(U_j)\right]}_{=:F_j(U)}dt\\
&&+\underbrace{\frac{g(U_j)}{\sqrt{\Delta x}}}_{G_{jj}(U)}dB_j,\\
\end{array}
\ee  
for $j=1,2,\ldots,N$ where $\{B_j\}_{k=1}^N$ are independent one-dimensional Brownian motions and the $N\times N$ matrix $L$ depends on 
the boundary conditions. For reflection conditions \cite{Gaines} it follows that $L_{jj}=-2$ if $j\in\{2,3,\ldots,N-1\}$, 
$L_{11}=-1=L_{NN}$, $L_{ij}=1$ if $|i-j|=1$ and $L_{ij}=0$ otherwise. For periodic conditions \cite{Gaines} one 
uses $L_{jj}=-2$ for all $j$, $L_{1N}=1=L_{N1}$, $L_{ij}=1$ if $|i-j|=1$ and $L_{ij}=0$ otherwise. For zero boundary
conditions \cite{Gyongy} the values $u_1^N\equiv 0\equiv u^N_N$ are fixed, the first and last equation in 
\eqref{eq:SODE} are discarded and the $(N-2)\times (N-2)$ matrix $L$ obeys $L_{jj}=-2$ for $j\in\{2,3,\ldots,N-1\}$, 
$L_{ij}=1$ if $|i-j|=1$ and $L_{ij}=0$ otherwise. For the simpler case \eqref{eq:1Dnoise} one has a single Brownian 
motion so that $B_j=B$ for all $j$ and the factor $1/\sqrt{\Delta x}$ in \eqref{eq:SODE} is removed \cite{Gaines}.

It remains to solve the SODE \eqref{eq:SODE} which can be more compactly written as 
\be
\label{eq:SODE1}
dU=F(U)dt+G(U)dB
\ee
where we view $F(U)=(F_1(U),\ldots,F_N(U))^T$, $B=(dB_1,\ldots,dB_N)^T$ as (column) vectors and $G(U)$ is a diagonal 
matrix with diagonal entries $G_{jj}(U)$. As a numerical scheme we shall always use either use the Euler-Maruyama method \cite{Higham} 
or the the Milstein method in its explicit \cite[p.345-351]{KloedenPlaten} or implicit \cite[p.399-404]{KloedenPlaten} form stated below.\\ 

\textit{Remark:} The Milstein method is usually good as an exploratory tool due to its robustness. 
It has strong order-one convergence \cite[Thm 10.3.5]{KloedenPlaten}. It is relatively straightforward to implement the Milstein method as no multiple 
stochastic integral evaluations occur since $x\in\R$ \cite[Chapter 10-11]{KloedenPlaten}\cite[p.2]{JentzenRoeckner}. 
Furthermore, it has recently been shown that it nicely extends to multiplicative trace-class noise 
\cite{JentzenRoeckner}. Hence, the Milstein method provides a quite remarkable compromise between theoretical error 
estimates and practical implementation issues; see also \cite{Higham} for a computational introduction with test
codes for scalar problems. The Euler-Maruyama method is faster but not as robust so it complements Milstein-methods nicely
if many sample paths have to be calculated for a well-understood parameter regime.\\  

To state both schemes consider 
$t\in[0,T]$ and define $(\Delta t):=T/K$ for some fixed $K\in\N$. Denote the numerical solution by $U^k\approx U(k(\Delta t))$ 
where $k\in\{0,1,2,\ldots,K\}$ and let $(\Delta B^k)=B((k+1)(\Delta t))-B(k(\Delta t))$ denote a vector of $\cN(0,\Delta t)$ normally 
distributed independent increments used at the $k$-th time step. The explicit Euler-Maruyama method is given by
\be
\label{eq:Euler}
U^{k+1}_j=U^k_j+\Delta t~F_j(U^k)+G_{jj}(U^k)~(\Delta B^k)_j
\ee
for each component $j\in\{1,2,\ldots,N\}$. The explicit Milstein scheme for \eqref{eq:SODE1} - with diagonal 
noise matrix $G$ - is given by \cite[p.348,(3.12)]{KloedenPlaten}
\be
\label{eq:Mils1}
\begin{array}{lcl}
U^{k+1}_j=U^k_j+\Delta t~F_j(U^k)+G_{jj}(U^k)~(\Delta B^k)_j\\
+\frac12 G_{jj}(U^k) 
\frac{\partial G_{jj}}{\partial U_j}(U^k)\left[(\Delta B^k)_j^2-\Delta t\right].\\
\end{array}
\ee
For non-diagonal noise satisfying a suitable commutativity condition the scheme 
is still quite simple \cite[p.348,(3.16)]{KloedenPlaten} while for more general cases one has to be careful \cite{JentzenRoeckner}.
The implicit version of the Milstein scheme for our problem is \cite[p.400]{KloedenPlaten}
\be
\label{eq:Mils2}
\begin{array}{lcl}
U^{k+1}_j&=&U^k_j+\Delta t~F_j(U^{k+1})+G_{jj}(\square)~(\Delta B^k)_j\\
&&+\frac12 G_{jj}(\square) 
\frac{\partial G_{jj}}{\partial U_j}(\square)\left[(\Delta B^k)_j^2-\Delta t\right]\\
\end{array}
\ee  
where we have the choice to make the scheme fully-implicit with $\square=U^{k+1}$ or semi-implicit with $\square=U^{k}$. Since 
the deterministic drift term $F$ causes the stability problems if $D (\Delta t)>(\Delta x)^2$ \cite[p.64,67]{Gaines} it makes 
sense to chose the semi-implicit version. The algebraic problem to solve for $U^{k+1}$ in \eqref{eq:Mils2} can be solved 
using standard techniques such as Newton's method. 

It should be noted that the convergence and error estimate of the numerical scheme do not immediately yield error estimates 
for quantitative properties or scaling laws of traveling waves for the FKPP equation. For example, it has been demonstrated 
\cite[p.71]{EbertvanSaarloos} that for pulled fronts of a discretized deterministic FKPP ($D=1=\mu$) the speed given to leading-order by
\be
\label{eq:wavespeed_discrete}
c^*=2-2(\Delta t)+\frac{1}{12}(\Delta x)^2+\cdots.
\ee  
A similar effect is expected for the stochastic FKPP equation and properties such as the diffusion properties of the 
wave speed. Therefore, we have to view numerical scaling laws as approximations which carry some information about the 
discretization step sizes $\Delta x$ and $\Delta t$. To minimize this effect, the formula \eqref{eq:wavespeed_discrete}, 
the stability requirement $\Delta t<(\Delta x)^2$ and the goal to minimize computation time indicate that we should
choose $\Delta t$ only slightly smaller than $(\Delta x)^2$. 

We are not interested in computing the exact wave speed, only its trend under parameter variation will be relevant here.
A simple method to compute an upper bound $\hat{c}$ on the wave speed for the initial condition $u(0,0)=1$ and $u(x,0)=0$ for
 a stochastic wave is to collect the set of points $(j~\Delta x,k ~\Delta t)$ such that $u(j~\Delta x,k ~\Delta t)$ is less 
and $u((j-1)~\Delta x,k ~\Delta t)$ is bigger than a threshold (usually we pick the threshold as $0.05$). For each point one
computes the estimate $c\approx \Delta x/\Delta t$ and obtains $\hat{c}$ as the maximum.  

\section{Linear Multiplicative Noise}
\label{sec:lin_noise}

\begin{figure*}[t]
\psfrag{x}{$x$}
\psfrag{t}{$t$}
\psfrag{u}{$u$}
\psfrag{a}{(a)}
\psfrag{b}{(b)}
\psfrag{c}{(c)}
  \includegraphics[width=1.0\textwidth]{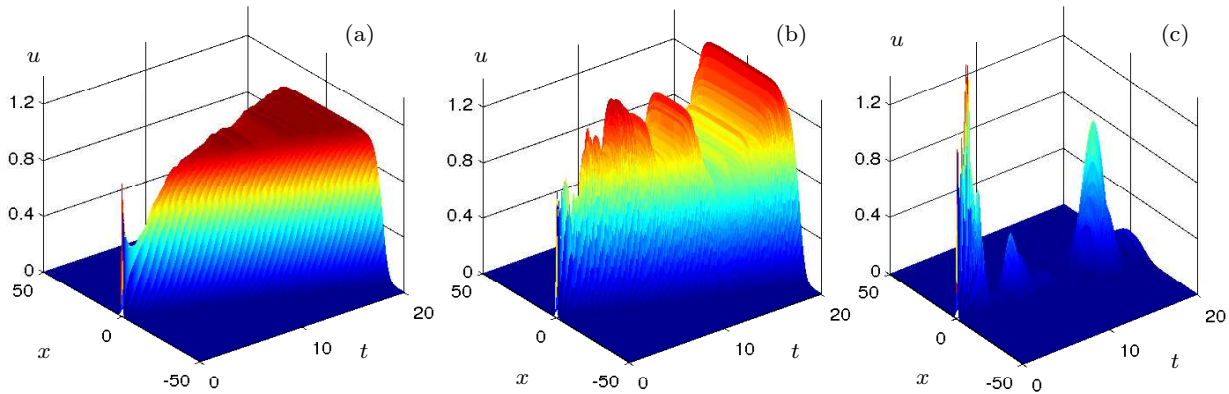}
\caption{Simulation of \eqref{eq:EZG1} using the implicit Milstein scheme \eqref{eq:Mils2} with parameters $K=100$, $T=20$, $N=10^3$ 
on the interval $[-50,50]$ with Neumann boundary conditions and initial condition $u(x,0)=1$ if $x=0$ and $u(x,0)=0$ otherwise. (a) 
$\epsilon=0.02$, (b) $\epsilon=0.3$ and (c) $\epsilon=1.2$.}
\label{fig:2}
\end{figure*}

The first stochastic version of the FKPP equation we consider was studied by Elworthy, Zhao and Gaines \cite{ElworthyZhaoGaines,Gaines} and is given 
by
\be
\label{eq:EZG}
\frac{\partial u}{\partial t}=\frac{\mu^2}{2}\frac{\partial^2u}{\partial x^2}+\frac{1}{\mu^2} u(1-u)+\hat{\epsilon} ~u~\dot{B}
\ee
with one-dimensional time-dependent white noise $\dot{B}$, a small parameter $0<\mu\ll1$ and noise strength $\hat{\epsilon}>0$. 
The multiplicative noise can be motivated {e.g.}~by the interaction of a population $u$ with the environment \cite{SpagnoloFiasconaroValenti} 
or by parameter noise in the deterministic part \cite[p.8]{RoccoRamirez-PiscinaCasademunt} such as a fluctuating growth rate \cite{VilarSole}. 
A multiplicative noise term of the form $g(u)=u$ has also been used in a model for plankton spreading \cite[eq.(4b)]{MalchowHilkerPetrovskiiBrauer}. 
For further mathematical background on SPDEs of the form \eqref{eq:EZG} we refer to Section \ref{sec:background_SPDE}. 

It is proven in \cite{ElworthyZhaoGaines} that there are three major regimes for \eqref{eq:EZG} 
depending upon the noise strength parameter $\hat{\epsilon}$. For some $\kappa=\cO(1)$ as $\mu\ra 0$, the cases $\hat{\epsilon}\sim\kappa/\mu^2$, 
$\hat{\epsilon}\sim \kappa/\mu$ and $\hat{\epsilon}\sim\kappa$ are identified as the strong, mild and weak noise regimes respectively. 
Elworthy, Zhao and Gaines prove and numerically demonstrate that for weak noise the wave propagation of the pulled front is 
basically unaffected while the wave fails to propagate in the strong noise regime \cite[Thms 8.1-8.3]{ElworthyZhaoGaines}. In the 
mild noise regime the wave speed is decreased as $\sqrt{2-\kappa}$ \cite[p.65]{Gaines}. It is important to note that the spontaneous
collapse of newly-introduced alien populations, which can occur in a strong noise regime, has been considered from an 
applied perspective in \cite{SimberloffGibbons}.   

Using the scaling law of Brownian motion and the transformation
\benn
x\mapsto \frac{x\mu^2}{\sqrt{2}},\qquad t\mapsto t\mu^2,\qquad \epsilon:=\mu\hat{\epsilon},
\eenn
as discussed in Section \ref{sec:background_FKPP}, in the SPDE \eqref{eq:EZG} yields the more familiar form of the FKPP equation
\be
\label{eq:EZG1}
\frac{\partial u}{\partial t}=\frac{\partial^2u}{\partial x^2}+u(1-u)+\epsilon ~u~\dot{B}.
\ee
This gives the quite natural view that $\epsilon\gg1$, $\epsilon\sim 1$ and $\epsilon\ll1$ are the strong, mild and weak noise regimes.
Figure \ref{fig:2} shows typical solutions for the three regimes; for details on the numerical methods see Section \ref{sec:background_numerics}. 
The initial condition is taken as the introduction of a species at a particular fixed location so that
\be
\label{eq:ic}
u(x,t=0)=\left\{\begin{array}{ll}
1 \quad & \text{if $x=0$,}\\
0 \quad & \text{otherwise.}\\
\end{array}\right.
\ee 
It is understood that the numerical initial condition is obtained by choosing a mesh having a mesh point $x=0$ with $u(0,0)=1$. 

Now consider the situation of the 'ecological observer' who can only measure the invasion wave at 
one particular point in space. Based on the results by 
Elworthy, Zhao and Gaines \cite{ElworthyZhaoGaines,Gaines} on propagation failure of the wave with 
increasing noise strength it is intuitive that the local statistics recorded at a 
fixed point carry information about the traveling front. 

\begin{figure}
\psfrag{u}{$\bar{U}$}
\psfrag{c}{$\hat{c}$}
\psfrag{u+}{$\bar{U}+\Sigma$}
\psfrag{u-}{$\bar{U}-\Sigma$}
\psfrag{epsilon}{$\epsilon$}
  \includegraphics[width=0.5\textwidth]{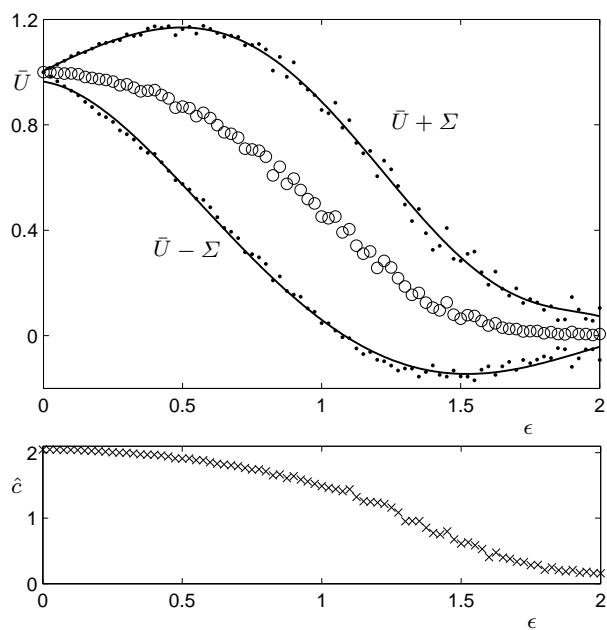}
\caption{Dependence of the time average $\bar{U}$ and the wave speed $c$ on the noise strength $\epsilon$ averaged 
over $200$ sample paths. 
The SPDE \eqref{eq:EZG1} has been numerically solved (using Euler-Maruyama \eqref{eq:Euler}) with 
$K=100$, $T=20$, $N=10^3$ on the interval $[-50,50]$ with Neumann boundary conditions and initial condition 
$u(x,0)=1$ if $x=0$ and $u(x,0)=0$ otherwise. The top part shows $\bar{U}$ (circles) which has been calculated 
as the mean of the time series $u(0,t)$ recorded by an ecological observer at the origin for $t\in[10,20]$. The dots indicate 
$\pm1$ standard deviation $\Sigma$ for the time series; the curves are associated interpolations forming a confidence 
neighborhood. The bottom part of the figure shows an (upper bound) estimate for the wave speed.}
\label{fig:1}
\end{figure}
   
For convenience we pick the point as $x=0$. Consider a single sample path 
$u(0,t)$. Let $T$ denote the final time and consider the two basic statistics
\benn
\begin{array}{lcl}
\bar{u}&=&\frac{1}{T-t_0}\int_{t_0}^T u(0,t)~dt,\\
\Sigma&=&\left[\frac{1}{T-t_0}\int_{t_0}^T (u(0,t)-\bar{u})^2 ~dt\right]^{1/2}.\\
\end{array}
\eenn
Figure \ref{fig:1} shows an average of $\bar{u}$ over $200$ sample paths which we denote by $\bar{U}$. From the 
results it is becoming clear, once one compares the $\bar{U}$ plot with the $\hat{c}$ plot, that a decreasing mean 
population size does provide the expected early-warning sign for a 
decreased invasion front speed. We also simulated the same case shown in Figure \ref{fig:1} for space-time 
white noise $\dot{W}$ in \eqref{eq:EZG1}. The results are qualitatively similar with a slight quantitative shift towards 
faster waves at comparable noise strength.

In both cases a relevant new result is that also the local fluctuations captured by the
variance show a quite interesting behavior. Consider the scenario where the actual carrying capacity for the 
population is unknown. In this case, the population level $\bar{u}$ is insufficient to determine how far we are 
from propagation failure of the wave. Naively, one may interpret small population fluctuations as an indicator for
a fast propagating wave but Figure \ref{fig:1} shows that it could equally well be a very slow propagating wave for 
a low population level. Hence one has to increase or decrease the noise strength to probe to which part of Figure \ref{fig:1}
the observations match.  

\section{Nonlinear Multiplicative Noise}
\label{sec:nonlin_noise}

As pointed out at the beginning of Section \ref{sec:lin_noise}, the noise terms $\epsilon u\dot{B}$ and $\epsilon u\dot{W}$
could be interpreted as parametric or environmental noise. Another possible source of noise is 'individual-based' or 
'finite-system-size' which we shall focus on in this section. M\"{u}ller and Tribe showed in \cite{MuellerTribe1} that the 
SPDE
\be
\label{eq:MT}
\frac{\partial u}{\partial t}=\frac16 \frac{\partial^2u}{\partial x^2}+\mu u(1-u)+\sqrt{2u}~\dot{W}
\ee 
arises as a limit of a contact process on a lattice originally studied in \cite{BramsonDurrettSwindle} as a model for 
long-range offspring displacement; traveling wave solutions 
to \eqref{eq:MT} exist for suitable parameter values \cite{Tribe}. However, M\"{u}ller and Tribe also studied the behavior of 
\eqref{eq:MT} with a scaled noise term $\sqrt{u}~\dot{W}$ varying the parameter $\mu$ and proved \cite[Thm 1]{MuellerTribe2} 
that there exists a critical value $\mu_c$, independent of $u(x,0)$, such that
\be
\label{eq:life_death}
\begin{array}{rcl}
\P(u(x,t)\text{ survives})&=&0 \text{ if $\mu<\mu_c$,}\\
\P(u(x,t)\text{ survives})&=&1 \text{ if $\mu>\mu_c$.}\\
\end{array}
\ee
Hence propagation failure of waves can occur like in the situation with noise term $u\dot{W}$. Using the mapping 
$t\mapsto t/\mu$, $x\mapsto x/\sqrt{6\mu}$, a standard scaling law result \cite[Lem 2.1.1]{MuellerTribe2} and the definition 
$\epsilon:=\sqrt{2}(6/\mu)^{1/4}$ transform \eqref{eq:MT} to
\be
\label{eq:MT1}
\frac{\partial u}{\partial t}=\frac{\partial^2u}{\partial x^2}+\mu u(1-u)+\epsilon\sqrt{u}~\dot{W}.
\ee

\begin{figure}
\psfrag{u}{$\bar{U}$}
\psfrag{c}{$\hat{c}$}
\psfrag{u+}{$\bar{U}+\Sigma$}
\psfrag{u-}{$\bar{U}-\Sigma$}
\psfrag{epsilon}{$\epsilon$}
  \includegraphics[width=0.5\textwidth]{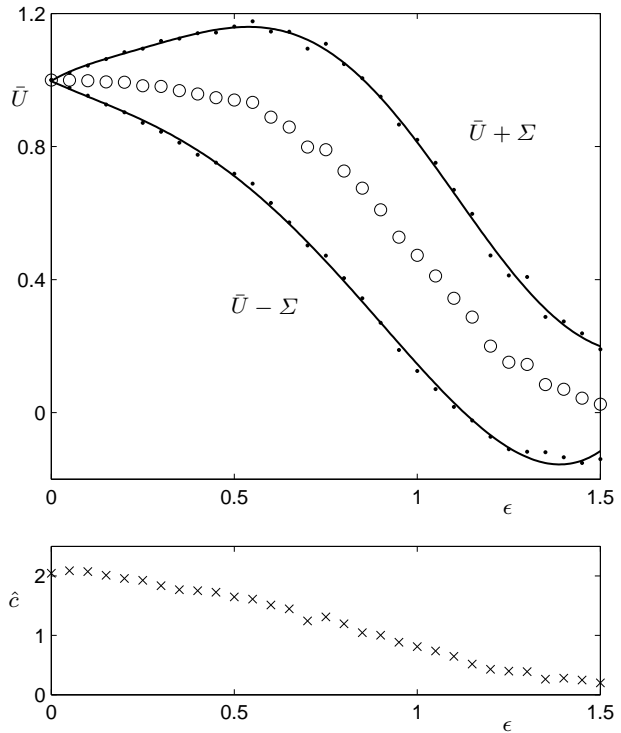}
\caption{Dependence of the time average $\bar{U}$ and the wave speed on the noise strength $\epsilon$ for \eqref{eq:MT1}.
Parameter values are as for Figure \ref{fig:1}.}
\label{fig:3}
\end{figure}
 
Figure \ref{fig:3} shows the dependence of the population level, its fluctuations and the wave speed on the 
parameter $\epsilon$. The results are very similar to Figure \ref{fig:1} with propagation failure for higher 
noise level as expected from \eqref{eq:life_death}. In particular, the conclusions from Section \ref{sec:lin_noise}
about inferring wave propagation properties from local data still apply. One may conjecture that the conclusions might 
apply to even more general versions of the FKPP equation with a noise term $\epsilon g(u)\dot{W}$ (or $\epsilon g(u)\dot{B}$) as long as $g(0)=0$.\\

However, the SPDEs \eqref{eq:EZG1} and \eqref{eq:MT1} have noise terms that increase monotonically with the 
population level. This may not be realistic in all situations as one expects the noise to change as $u$ approaches 
the carrying capacity. This is one motivation to study the SPDE
 \be
\label{eq:MS}
\frac{\partial u}{\partial t}=\frac{\partial^2u}{\partial x^2}+\mu u(1-u)+\epsilon\sqrt{u(1-u)}~\dot{W}.
\ee
This model was studied by several groups. In \cite{MuellerSowers} it was proved for 
sufficiently small noise that compactly supported initial data remain within a time-dependent 
interval and that a well-defined front as well as an asymptotic wave speed exist; interestingly, the shape and asymptotic
form of waves has been of interest by an independent group for a discrete stochastic model \cite{LewisPacala}. 

Detailed numerical studies of the wave speed for \eqref{eq:MS} have been carried out \cite{PechenikLevine,BrunetDerridaMuellerMunier} 
focusing on the small
noise regime and the fluctuation properties of the front. Due to the special structure of the FKPP equation one
may also exploit a duality argument of \eqref{eq:MS} to a particle system \cite{DoeringMuellerSmereka,ShigaUchiyama}. 
Doering, M\"{u}ller and Smereka conjecture \cite[eq (55)]{DoeringMuellerSmereka} from the duality relation 
that the asymptotic wave speed in the strong noise regime is given by $c\sim 2/\epsilon^2$ as $\epsilon\ra \I$. Although 
it is unclear whether this conjecture is correct it is evident from numerical simulations \cite[Fig 2]{DoeringMuellerSmereka} 
that the wave speed decreases upon increasing the noise.  

\begin{figure}
\psfrag{u}{$\bar{U}$}
\psfrag{c}{$\hat{c}$}
\psfrag{u+}{$\bar{U}+\Sigma$}
\psfrag{u-}{$\bar{U}-\Sigma$}
\psfrag{epsilon}{$\epsilon$}
  \includegraphics[width=0.5\textwidth]{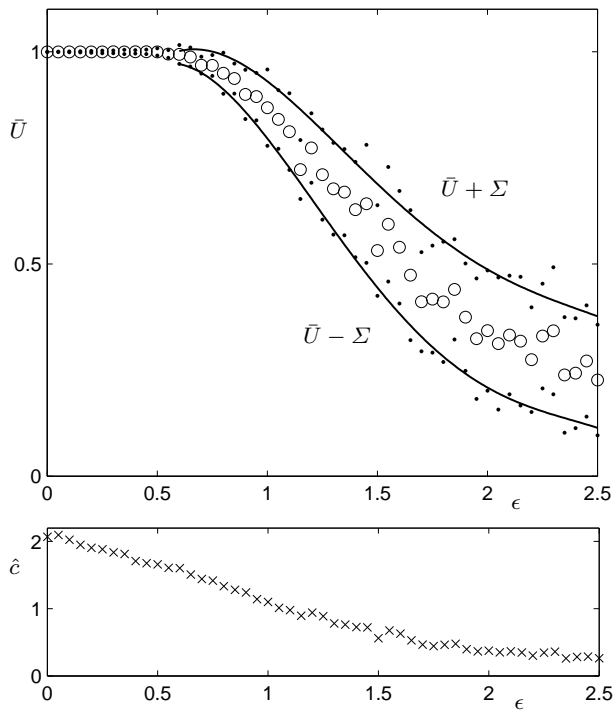}
\caption{Dependence of the time average $\bar{U}$ and the wave speed on the noise strength $\epsilon$ for \eqref{eq:MS}.
Parameter values are as for Figure \ref{fig:1} except for the slightly smaller time-step size $N=3\cdot 10^3$.}
\label{fig:4}
\end{figure}

Figure \ref{fig:4} shows the mean, standard deviation and wave speed calculated for \eqref{eq:MS}. There are some minor 
differences between this case and $g(u)=u$ and $g(u)=\sqrt{u}$ shown in Figures \ref{fig:1} and \ref{fig:3}. There is a 
larger plateau for small noise and it takes larger noise strengths to reach the vicinity of propagation failure. However,
the main warning-signs from local data still remain as it is still possible to conclude from large population levels 
and low fluctuations a fast wave while increasing fluctuations lead to slower speeds and low population levels with
smaller fluctuations indicate closeness to propagation failure. 

Another different form of the noise term given by $g(u)=u(1-u)$ was considered in \cite[eq (39)-(41)]{RoccoRamirez-PiscinaCasademunt}
but we shall not consider it here as the results are similar. 

In summary, one should always measure the closeness to carrying capacity and the size of the fluctuations (standard deviation, variance).
If system parameters change slowly one may determine from Figures \ref{fig:1}-\ref{fig:4} whether the distance to propagation failure 
has increased or decreased.

\section{Transitions for the Allee Effect}
\label{sec:Allee}

Although propagation failure is extremely interesting from the viewpoint of critical transitions, it is certainly not
the only invasion wave phenomenon where local early-warning signs are desirable. As already discussed in Section 
\ref{sec:background_FKPP} there can also be pushed fronts if the nonlinearity $f(u)$ is chosen differently. A 
reasonable prototypical model to study is the following SPDE 
\be
\label{eq:RRC}
\frac{\partial u}{\partial t}=\frac{\partial^2u}{\partial x^2}+u(1-u)(u-\mu)+\epsilon g(u)~\dot{W}
\ee 
which has been considered in \cite{RoccoRamirez-PiscinaCasademunt}. The nonlinearity $f(u)=u(1-u)(u+\mu)$ may 
obviously arise due to an Allee effect in the context of ecology but it is also commonly used in other areas
of mathematical biology, {e.g.}~in neuroscience \eqref{eq:RRC} would be referred to as Nagumo's equation \cite{Nagumo}.
We briefly recall some results about the deterministic PDE ($\epsilon=0$) described in \cite{Popovic}. For 
$\mu\in[-1/2,1/2]$ there exists a closed-form wave
\benn
u(x,t)=u(x-ct)=u(\xi)=\frac{1}{1+e^{\frac{1}{\sqrt{2}}(\xi-\xi^-)}}
\eenn
for an arbitrary phase $\xi^->0$ and propagation speed 
\be
\label{eq:wavespeed_Nagumo}
c=\frac{1}{\sqrt{2}}-\sqrt{2}\mu.
\ee
There are three interesting special points. For $\mu=1/2$ the front speed vanishes and the front reverses direction if $\mu>1/2$. The regime for 
$\mu\in(0,1/2)$ is called bistable and changes to a pushed front at $\mu=0$; see \cite[Sec 1]{Popovic} and references 
therein for details. The pushed front regime applies for $\mu\in(-1/2,0)$ and at $\mu=-1/2$ there is a pushed-to-pulled front 
transition \cite[Sec V]{RoccoRamirez-PiscinaCasademunt}.
The wave speed for the pulled front is $c^*=2\sqrt{-\mu}$ \cite[Sec 3.3]{Popovic1}\cite[p.38-42]{vanSaarloos}. Therefore,
it is interesting to try to find early-warning signs for approaching the three special 
points $\mu=-1/2,0,1/2$ cases. The front reversal case $\mu= 1/2$ is clearly important as a direction change for an 
invasion front could be regarded as a critical transition but the other two cases could be of interest as well.  

\begin{center}
\begin{figure}
\psfrag{u}{$\bar{U}$}
\psfrag{mu}{$\mu$}
\psfrag{u+}{$\bar{U}+\Sigma$}
\psfrag{u-}{$\bar{U}-\Sigma$}
  \includegraphics[width=0.4\textwidth]{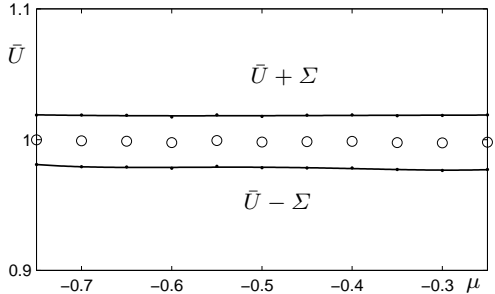}
\caption{Dependence of the time average $\bar{U}$ on the noise strength $\epsilon$ averaged 
over $200$ sample paths. 
The SPDE \eqref{eq:RRC} for $g(u)=u$ has been numerically solved (using Euler-Maruyama \eqref{eq:Euler}) with 
$K=100$, $T=15$, $N=10^3$ on the interval $[-50,50]$ with Neumann boundary conditions and initial condition 
$u(x,0)=1$ if $x\in[-1,1]$ and $u(x,0)=0$ otherwise. $\bar{U}$ (circles) has been calculated 
as the mean of the time series $u(0,t)$ recorded by an ecological observer at the origin for $t\in[7.5,15]$. The dots indicate 
$\pm1$ standard deviation $\Sigma$ for the time series; the curves are associated interpolations forming a confidence 
neighborhood.}
\label{fig:5}
\end{figure}
\end{center}

For the SPDE \eqref{eq:RRC} we shall choose the simple multiplicative noise $g(u)=u$. We consider the pushed-to-pulled
transition first and try to apply our approach from Sections \ref{sec:lin_noise}-\ref{sec:nonlin_noise}. Figure \ref{fig:5}
shows the analog of the top parts of Figures \ref{fig:1}-\ref{fig:4}. We observe that it is impossible to detect a trend or infer 
the speed of the wave. Hence, for the pushed-to-pulled transition with space-time white noise the classical 
variance-based early-warning signs cannot be applied for local data perturbed by a fixed noise level and observed 
at the center of the wave. Therefore, one should also think of new early-warning sign techniques in the 
context of wave propagation. 

\begin{figure*}
\psfrag{x}{$x$}
\psfrag{t}{$t$}
\psfrag{u}{$u$}
\psfrag{a}{(a)}
\psfrag{b}{(b)}
\psfrag{c}{(c)}
  \includegraphics[width=1.0\textwidth]{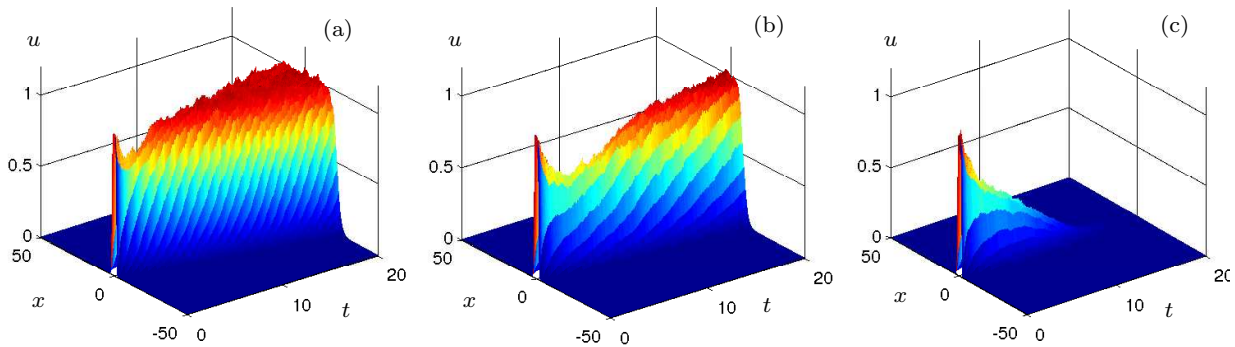}
\caption{Simulation of \eqref{eq:RRC} for $g(u)=u$ using the implicit Milstein scheme \eqref{eq:Mils2} with parameters $K=100$, $T=20$, $N=2\cdot10^3$ 
on the interval $[-50,50]$ with Neumann boundary conditions and initial condition $u(x,0)=1$ if $x\in[-1,1]$ and $u(x,0)=0$ otherwise. (a) 
$\mu=-0.3$, (b) $\mu=0.2$ and (c) $\mu=0.4$.}
\label{fig:6}
\end{figure*}

Note carefully that we always used in our computations in Figures \ref{fig:1}-\ref{fig:5} the regime for $u(0,t)$ when the 
wave is already fully formed with $t\in[T_0,T]$ for some $T_0\gg 1$. However, the transient regime starting from the 
localized initial condition may also contain important information. Figure \ref{fig:6} shows three numerical simulations 
for $\mu=-0.3,0.2,0.4$. 

\begin{figure}
\psfrag{umin}{$u_{m}$}
\psfrag{mu}{$\mu$}
\psfrag{a}{(a)}
\psfrag{b}{(b)}
\psfrag{c}{(c)}
  \includegraphics[width=0.5\textwidth]{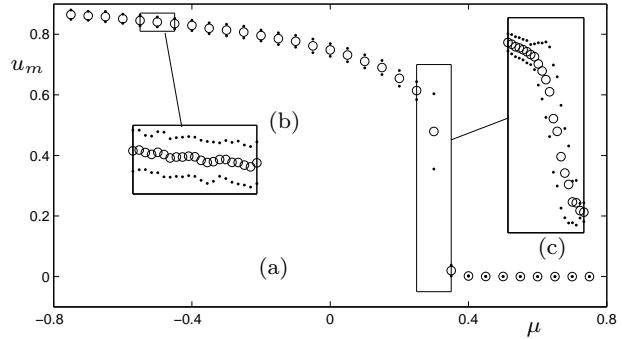}
\caption{(a) Dependence of the minimum $u_{m}$ defined in \ref{eq:umin} on $\mu$ with $T_0=10$, averaged 
over $200$ sample paths. 
The SPDE \eqref{eq:RRC} for $g(u)=u$ has been numerically solved (using Euler-Maruyama \eqref{eq:Euler}) with 
$K=100$, $T=20$, $\epsilon=0.05$, $N=10^3$ on the interval $[-50,50]$ with Neumann boundary conditions and initial condition 
$u(x,0)=1$ if $x\in[-1,1]$ and $u(x,0)=0$ otherwise. The circles indicate $u_{m}$ and the dots
$\pm1$ standard deviation $\Sigma$ calculated from the sample paths. (b) Zoom near the theoretical pushed-to-pulled 
transition at $\mu=1/2$. (c) Zoom near propagation failure transition.}
\label{fig:7}
\end{figure}

The computation suggests that the initial
transient spreading of the wave $u(0,t)$ for $t\in[0,T_0]$ is interesting. A simple measure to consider is
\be
\label{eq:umin}
u_{m}:=\min_{t\in [0,T_0]}\{u(0,t): \text{ for a given $u(x,0)$}\}.
\ee
Clearly, the result depends upon the choice of $T_0$ and the initial condition $u(x,0)$. However, if both are fixed then we
may compare the results. Figure \ref{fig:7}(a) shows the results for a parametric study of $\mu\in[-0.75,0.5]$. The insets (b)-(c) show a 
finer mesh resolution near the pushed-to-pulled transition at $\mu= -1/2$ and near propagation failure which is slightly
shifted from the theoretical value at $\mu=1/2$ as the small finite-width initial condition and the noise both seem to 
contribute to reach the absorbing state $u\equiv 0$ for parameter values $\mu$ smaller than $1/2$. In fact, due to these 
effects, the transition is more drastic than the formula \eqref{eq:wavespeed_Nagumo} predicts. 

From Figure \ref{fig:7}(b) it is apparent that the pushed-to-pulled transition is probably unpredictable from local data collected 
at $x=0$. Since the wave speed transition is continuous one should probably not classify the pushed-to-pulled transition 
as a 'critical transition'. Therefore, it is not crucial to predict it but the result shows the limitation of the ecological 
observer at $x=0$. The same conclusion applies to the change from the pushed to the bistable regime at $\mu=0$. For the 
propagation failure scenario Figure \ref{fig:7}(c) shows a scaling law for the decrease of $u_{m}$ and a slightly increasing 
variance may help us to anticipate the upcoming critical transition. This should not be surprising since 
we already considered similar propagation failure cases in Sections \ref{sec:lin_noise}-\ref{sec:nonlin_noise}. The 
difference is that we used a completely different indicator in Figure \ref{fig:7}(c). 

As for the classical FKPP equation one should remark for the Allee effect situation that different noise terms certainly do 
make sense, {e.g.}~$g(u)=u(1-u)$ considered in \cite[eq.(2)]{Armeroetal}. Based on the observations for varying the noise 
terms for the classical FKPP equation, and obtaining similar results for several choices, we shall not consider these
generalizations for \eqref{eq:RRC}.
 
\section{Noncompact Initial Invasions}
\label{sec:noncompact}

Based on the results in Section \ref{sec:Allee} we have observed that the initial transient regime, starting from a localized invasion 
wave, can be useful. It remains to consider the case when the initial condition is not localized. In particular, we consider 
the FKPP equation 
\be
\label{eq:IC1}
\frac{\partial u}{\partial t}=\frac{\partial^2u}{\partial x^2}+u(1-u)+\epsilon ~u~\dot{W}.
\ee
with initial condition
\be
\label{eq:IC2}
u(x,0)=e^{-\alpha |x|},\quad \text{for $\alpha>0$.}
\ee
Recall from Section \ref{sec:background_FKPP} that the wave speed scales as $c(\alpha)=\cO(1/\alpha)$ for $\alpha \ra 0$. 

\begin{figure}
\psfrag{um}{$u_{m}$}
\psfrag{mu}{$\mu$}
\psfrag{a}{(a)}
\psfrag{b}{(b)}
\psfrag{c}{$\hat{c}$}
  \includegraphics[width=0.5\textwidth]{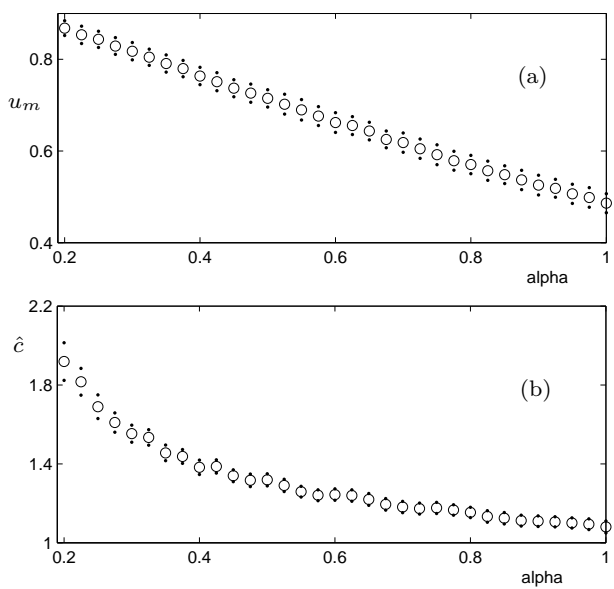}
\caption{Dependence of the minimum $u_{m}$ defined in \ref{eq:umin} on $\alpha$ for \eqref{eq:IC1}-\eqref{eq:IC2}; average 
over $200$ sample paths. 
The parameters for the numerical simulation(using Euler-Maruyama \eqref{eq:Euler}) on the interval $[-50,50]$ with 
Neumann boundary conditions are $K=150$, $T=10$, $\epsilon=0.05$ and $N=500$. (a) The circles indicate $u_{m}$ and the dots
$\pm1$ standard deviation calculated from the sample paths. (b) Wave speed $\hat{c}$ (circles) and associated $\pm 1$ standard deviation (dots).}
\label{fig:8}
\end{figure}

Figure \ref{fig:8}(a) shows the dependence of the initial transient observed at $x=0$ on the complete initial data. Therefore,
small minimum values indicate comparatively slower waves and no response to the initial condition ($u_m\approx 1$) signals a 
very fast wave. Figure \ref{fig:8}(b) shows an upper bound to the wave speed and raises the interesting question whether we should,
or should not, view a blow-up point for the wave speed as a critical transition.

There are two main conclusions from the results for \eqref{eq:IC1}-\eqref{eq:IC2} and from Section \ref{sec:Allee}. Firstly, one 
definitely should try to measure an invasion wave immediately once the first occurrence of a new population in a new environment
has been observed. Secondly, knowing the basic structure of the initial condition can be crucial for prediction, {e.g.}~in Figure \ref{fig:7}   
$0\ll u_m<1$ still indicates a well-defined asymptotic wave speed in the pushed regime while for Figure \ref{fig:8} the condition $0\ll u_m<1$
indicates closeness to a wave speed blow-up point. Hence, it is crucial to know, on a qualitative level, whether the initial invasion is
really localized or whether it really consists of a full front. 

\section{Outlook}
\label{sec:outlook}

Since early-warning signs and stochastic scaling laws for noisy traveling waves are still a relatively new direction, we 
have only been able to cover a few aspects here. Many open problems arose which we summarize here.

The restrictions to one spatial dimension $x \in\R$ and one population component $u$ have to removed in the future. There are many interesting
cases {e.g.}~multi-component systems such as reaction-diffusion models with predation \cite{OwenLewis}, FKPP-type plankton dynamics 
\cite{BrindleyBiktashevTsyganov} or Nagumo (Allee effect)-type equations \cite{GuckenheimerKuehn3}. Multiple spatial dimensions can lead to more
complicated bifurcation structures \cite{JordanPuri}. One may also remove all restrictions which can generate interesting life-death transitions 
for multi-component, 2D and 3D systems \cite{MuellerTribe}. Another highly relevant generalization are heterogeneous \cite{RoquesHamel} and 
random \cite{Xin} environments. Furthermore, the structure of the FKPP equation may be too restrictive which suggests to add transport/advection 
terms and active boundaries in which case discontinuous wave speed transitions have been reported \cite{CostaBlytheEvans}. Also the 
assumption of time-white or space-time-white noise is too restrictive and one should extend the view to spatially-colored noise \cite{GarciaOjalvoSancho}
and trace-class covariance operators. Another issue that looked interesting is the relevance of fluctuations ('front diffusion') 
\cite{ArmeroCasademuntRamirez-PiscinaSancho,RoccoCasademuntEbertvanSaarloos} for early-warning signs.

In all cases, our main driving question in this paper seems to be open: How much information do local statistics of an SPDE, collected
at one (or multiple) locations, carry about the speed and bifurcations of traveling waves? It is seems plausible to obtain basic answers to these 
questions using numerical simulations. To develop a mathematical theory for quantitative scaling laws of SPDEs and their application to 
critical transitions is expected to be a challenging problem for a long time.\\ 

\textbf{Acknowledgements:} I would like to thank the European Commission (EC/REA) for support by a Marie-Curie International Re-integration Grant.

\bibliographystyle{plain}
\bibliography{../my_refs}

\end{document}